\documentclass[useAMS,usenatbib]{mn2e}
\usepackage{graphicx}
\usepackage{amsmath,amsfonts,amssymb}
\usepackage{wrapfig}          
\usepackage{graphicx}         
\usepackage{epsfig}
\usepackage{psfrag}
\usepackage{pstricks,pst-plot} 

\begin{document}

\title[binding energy of neutron stars]
{The role of binding energies of neutron stars on the accretion driven evolution}

\author[Bagchi]
{\parbox[t]{\textwidth}{Manjari Bagchi$^{1}$}\\
\vspace*{3pt} \\
\\ $^1$ Physics Department, West Virginia University, Morgantown, WV 26506}
\maketitle
\begin{abstract}
Millisecond pulsars are believed to descend from low mass x-ray binaries. Observable parameters of binary millisecond pulsars \textit{e.g.} mass of the pulsar, mass of the companion, spin period of the pulsar, orbital period, orbital eccentricity etc are used to probe the past accretion history of the millisecond pulsars. But unfortunately in these studies, binding energy of the neutron star is not considered commonly. We show that the effect of the binding energy is significant in the estimation of the amount of mass accretion and thus should be incorporated in models for binary evolutions. Moreover, different Equations of State for dense matter give different values for the accreted mass for the same amount of increase in the gravitational mass of the neutron star implying the need of constraining dense matter Equations of State even to understand the spin-up procedure properly.
\end{abstract}


\section{Introduction} 
 
Millisecond radio pulsars (MSRPs) are believed to descend from the neutron stars in low mass X-ray binaries (LMXBs) that have been spun up by acquiring angular momentum through accretion of matter from their companions \citep{alpar82, rs82}. kHz quasiperiodic oscillations and millisecond pulse periods of neutron stars in several LMXBs support this theory (\citet{wv98} and references therein) and the most compelling proof came from the MSRP J1023+0038 which has lost its accretion disc within the past decade \citep{arch09}. In spite of continuing efforts to understand the evolution of LMXBs \citep{wrs83, rvj83, ps88, bvdh91, ity95, prp03, ndm04, bkr08}, sometimes discoveries of MSRPs with peculiar characteristics impose problems and compel us to think of alternative scenarios. Two such systems are binary MSRPs PSR J1903+0327 which has an eccentric orbit \citep{crl08} and PSR J1614-2230 which is highly recycled with comparatively larger companion mass \citep{dp10}; and special evolutionary channels have been proposed for these objects \citep{fbw10, lin10}. But all these evolutionary models usually ignore the effects of binding energies of neutron stars which should be taken care of for a more realistic understanding of the binary evolution as neutron stars posses significant amounts of binding energies. In the present work, we show that the change in the total binding energy in accretion driven evolution is significant and should be incorporated in realistic models of binary evolutions.

\section{Study of binding energy of neutron stars} 
\label{sec:work}

The gravitational mass of a neutron star is defined as $M_G = \int_{0}^{R} 4 \pi r^2 \rho(r) \, dr$ where $\rho(r)$ is the mass density at a radial distance $r$ from the center and $R$ is the radius of the neutron star. This is the measurable entity and being used in studies of neutron star mass distributions \citep{tc99,spr10,kkt10}. But in the case of relativistic objects like neutron stars, one should think about the proper volume element $d\cal{V}$ which becomes $4 \pi r^2 \left[1-2 G m_{_G}(r)/rc^2 \right]^{-1/2} \, dr$ in Schwarzschild geometry. So the proper mass of the neutron star becomes $ M_P = \int_{0}^{R} 4 \pi r^2 \left[1-2 G m_{_G}(r)/rc^2 \right]^{-1/2} \rho(r) \, dr$ and the total number of particles (baryons) becomes $ N_B= \int_{0}^{R} 4 \pi r^2 \left[1-2 G m_{_G}(r)/rc^2 \right]^{-1/2} n(r) \, dr$ where $ n(r)$ is the particle number density at $r$. This makes one define the baryonic mass as $M_B= m_B \, \int_{0}^{R} 4 \pi r^2 \left[1-2 G m_{_G}(r)/rc^2 \right]^{-1/2} n(r) \, dr$ where $m_B$ is the mass of a baryon. The binding energy of the neutron star is $BE = (M_B - M_G) \, c^2$ where $c$ is the velocity of light and the gravitational binding energy of the neutron star is $BE_G= (M_P - M_G) \, c^2$. 
As the only difference in the expression of $M_P$ from that of $M_G$ is the appearance of the metric element $\left[1-2 G m_{_G}(r)/rc^2 \right]^{-1/2}$ inside the integral, so $BE_G$ is actually the manifestation of the space-time curvature for objects as compact as neutron stars. Solving Tollman-Oppenheimer-Volkov (TOV) equations of hydrostatic equilibrium, one can obtain the values of all these entities with a suitable choice of the Equation of State (EoS). It is noteworthy to mention here that although TOV equations have been derived for non-rotating stars in Schwarzschild geometry, these are good enough even for the millisecond pulsars. The definition of the proper volume $d\cal{V}$ is different for a rotating-star where the metric is different, but we have found with the freely available ``Rapidly Rotating Neutron Star" code by Nikolaos Stergioulas\footnote{http://www.gravity.phys.uwm.edu/rns/} that for a fixed value of $M_{G}$, $M_{P}$ usually decreases by an amount less than 1\% if we increase the angular velocity $\omega$ from zero to 3000 $s^{-1}$. So in the present work, we keep ourself confined within Schwarzschild geometry. One important question that arises here is what is the value of $m_B$ ? Following \citet{ab77}, we set $m_B$ equal to one atomic mass unit ``\textit{u}" (based on $C^{12}$ scale, 931.49402 MeV) which is appropriate as pre-collapse evolved stars contain large amounts of $C^{12}$ or nuclei of similar binding energy per nucleon. The mass of $H^{1}$ is 1.007825 \textit{u} and the mass of $He^{4}$ is 4.002602 \textit{u}, so choosing $m_B = 1$ \textit{u} is valid even for the present work although we are mainly concerned about the change in $M_B$ due to the accretion of hydrogen or helium rich matter on the neutron star from its binary companion (usually in the red-giant phase).

Among numerous nuclear EsoS, we choose a standard EoS APR \citep{apr98} which agrees with the possible $M_G - R$ regions for three neutron stars determined from X-ray observations of thermonuclear bursts \citep{ozel10} as well as gives maximum value of $M_G$ to be greater than the mass of the most massive neutron star discovered till date - the millisecond pulsar PSR J1614-2230 having $M_G = (1.97 \pm 0.04)~M_{\odot}$ \citep{dp10}. But there is a possibility of discovering even heavier neutron stars in future - the ``black widow" pulsar B1957+20 may be one such example which has the lower limit of $M_G$ as $1.66~M_{\odot}$ with the best fit value as $2.40 \pm 0.12~M_{\odot}$ \citep{kerk10}. Further study is needed to get a precise mass measurement for this neutron star. But at the present situation, one should not exclude very stiff EsoS in any study involving EsoS. As an example of such a stiff EoS, we use MS0 \citep{ms96} although it disagrees with the findings of \citet{ozel10}. As an EoS much softer that APR would not give $M_G$ as high as $1.97~M_{\odot}$, we do not use any such EoS. It is worthwhile to mention here that the statistical method of \citet{slb10} favors EsoS which are even softer than APR at high densities (see their figure 8). But their results fail to give $M_G$ as high as $2.4~M_{\odot}$ and there is enough scope to improve their method (see their discussions). On the other hand, the study of the cooling phase of the X-ray burster 4U 1724-307 by \citet{sul10} favors an EoS as stiff as MS0.

\begin{figure}
\centerline{\psfig{figure=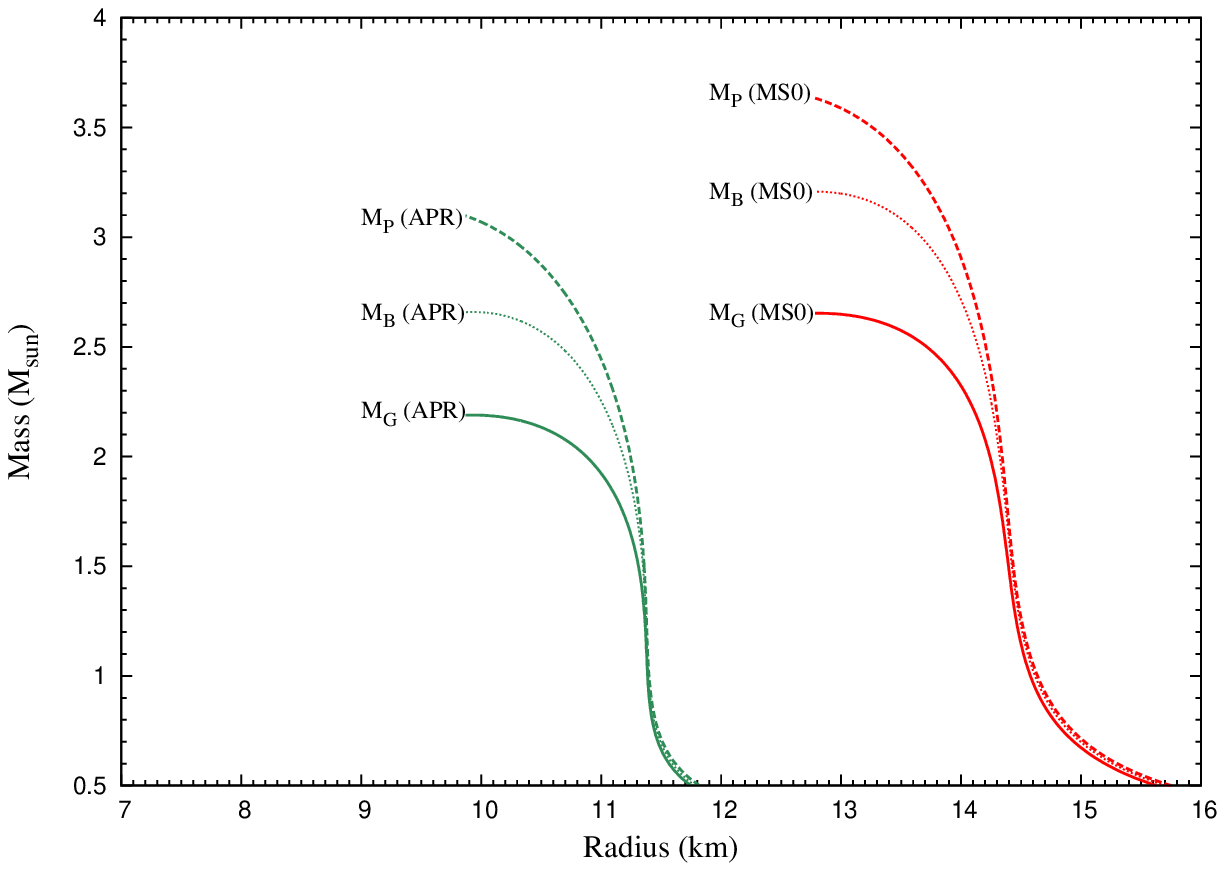,width=7cm}}
\centerline{\psfig{figure=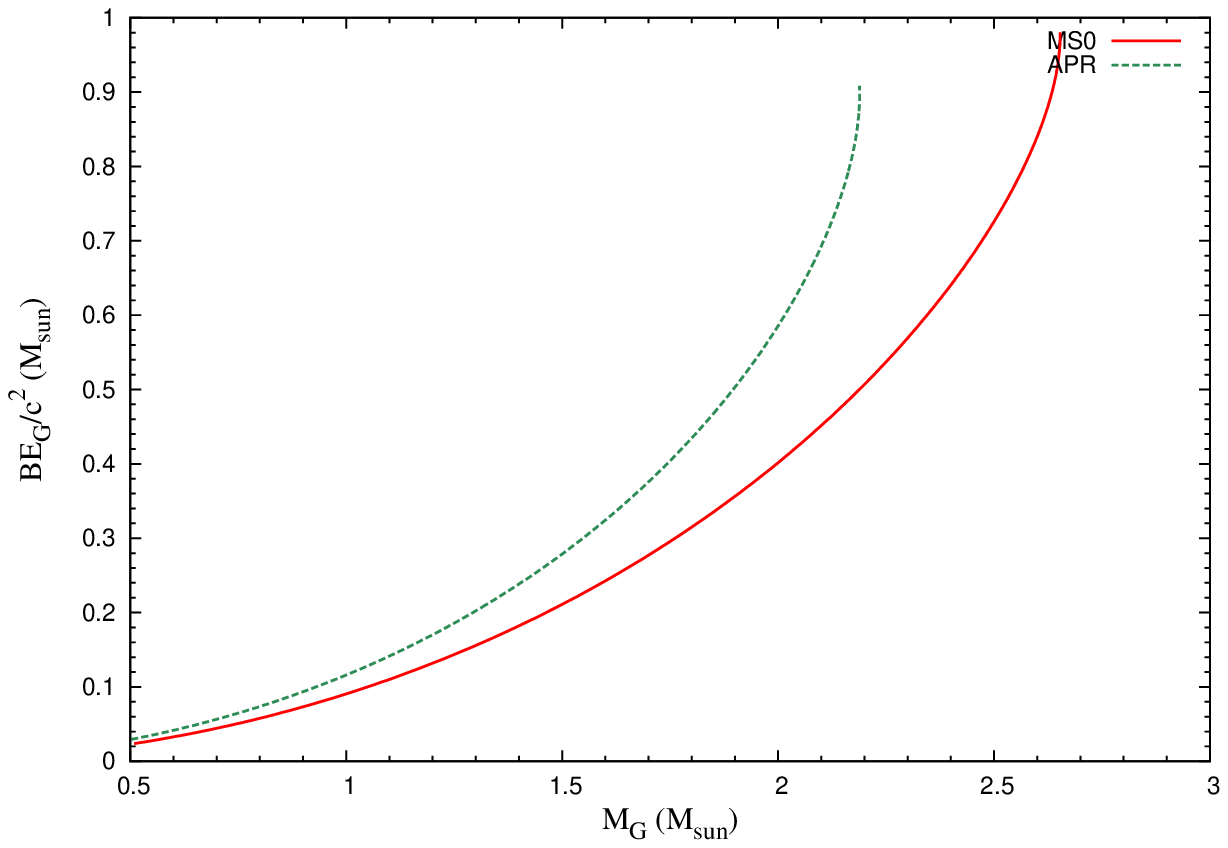,width=7cm}}
\centerline{\psfig{figure=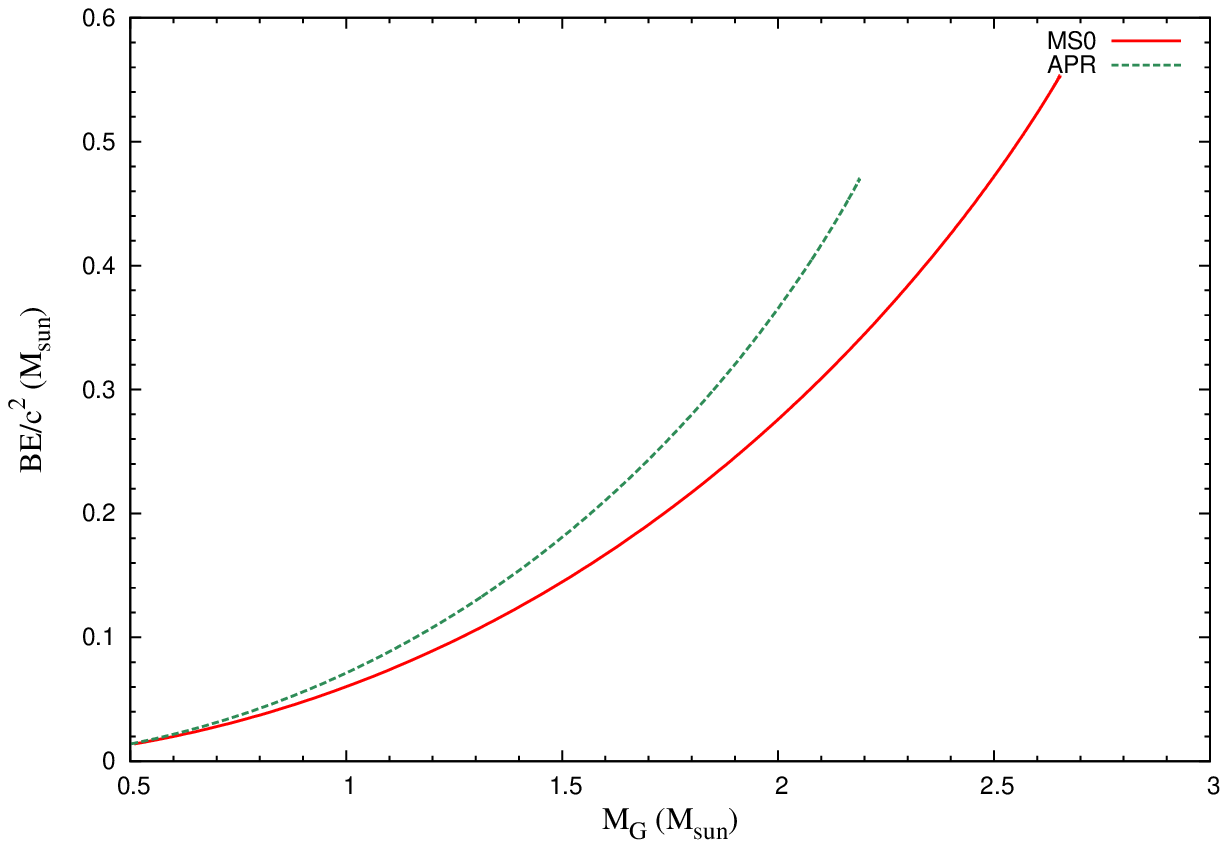,width=7cm}}
\caption{The upper panel shows Mass-Radius curves for neutron stars as obtained using EoS MS0 and EoS APR. The values of $M_B$ have been computed using $m_B = 1~u$. The middle panel shows the variations of the gravitational binding energy $BE_G$ with $M_G$ and the bottom panel shows the variations of binding energy $BE$ with $M_G$ for the same set of EsoS. } \label{fig:mrbn}
\end{figure}

In the top panel of Fig. \ref{fig:mrbn}, we show mass-radius relations of neutron stars as obtained by solving TOV equations with these two EsoS. It is clear that the difference between three types of masses increases with the increase of $M_{G}$ leading to the increase in $BE_G$ and $BE$ with $M_G$ (shown in the middle and bottom panels of the same figure). This happens because with the increase of $M_G$, neutron stars become more compact and the general relativistic effects more dominant. For a fixed value of $M_G$, the values of $BE_G$ and $BE$ are higher for a softer EoS than those for a stiffer EoS due to the higher compactness of the neutron star in the first case. This fact had been observed earlier by \citet{pod05}. While $BE_G$ is useful to understand general relativistic effects, it is $BE$ which is relevant in case of the accretion driven evolution which we discuss next.

We study the role of the binding energy in the evolution of millisecond pulsars which are the results of the mass accretion onto the initial slow neutron stars from their companions. Let us assume a millisecond pulsar with gravitational mass $M_{G,f}$ and binding energy $BE_{f}$ initially was a slow neutron star with the gravitational mass $M_{G,i}$ and binding energy $BE_{i}$ before the accretion process started. Here $M_{G,f} > M_{G,i}$ implying $BE_{f} > BE_{i}$ (bottom panel of Fig. \ref{fig:mrbn}). The increase in the gravitational mass is $\Delta M_{G} = M_{G,f} - M_{G,i}$ and the increase in the binding energy is $\Delta BE = BE_{f} - BE_{i} $ and the total amount of mass accreted is given by $M_{acc} = \Delta M_{G} + \Delta BE / c^2 = \Delta M_{B}$. Unfortunately, people usually assume $M_{acc} = \Delta M_{G}$; one of the main aims of the present paper is to point out this error. 

\begin{figure}
\centerline{\psfig{figure=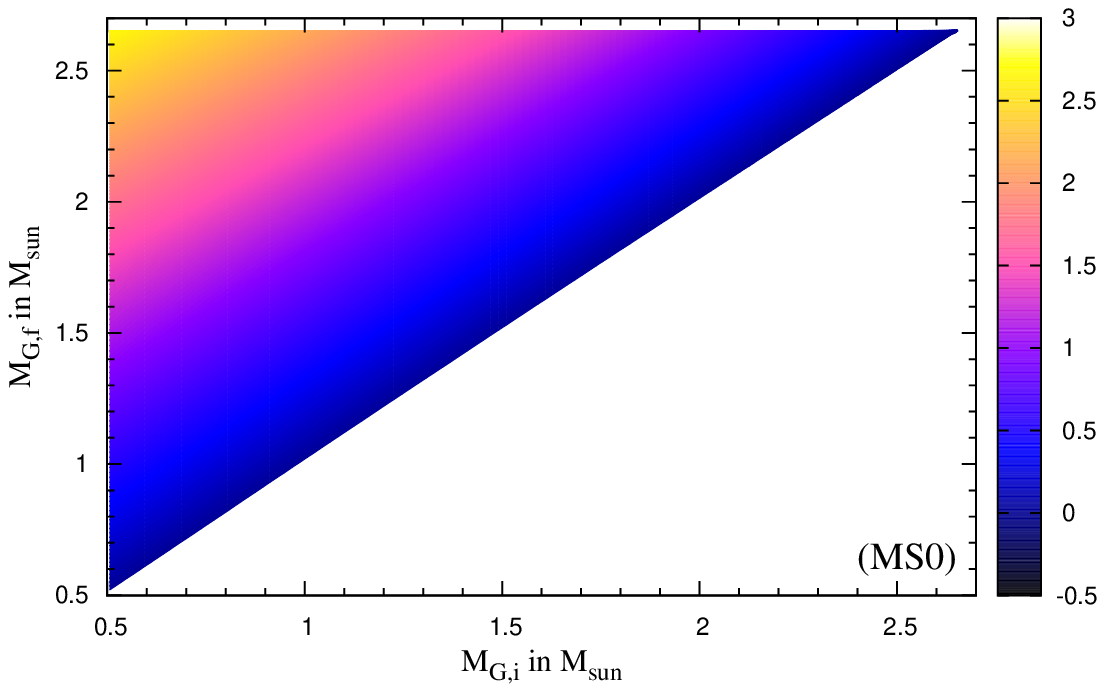,width=10cm}}
\centerline{\psfig{figure=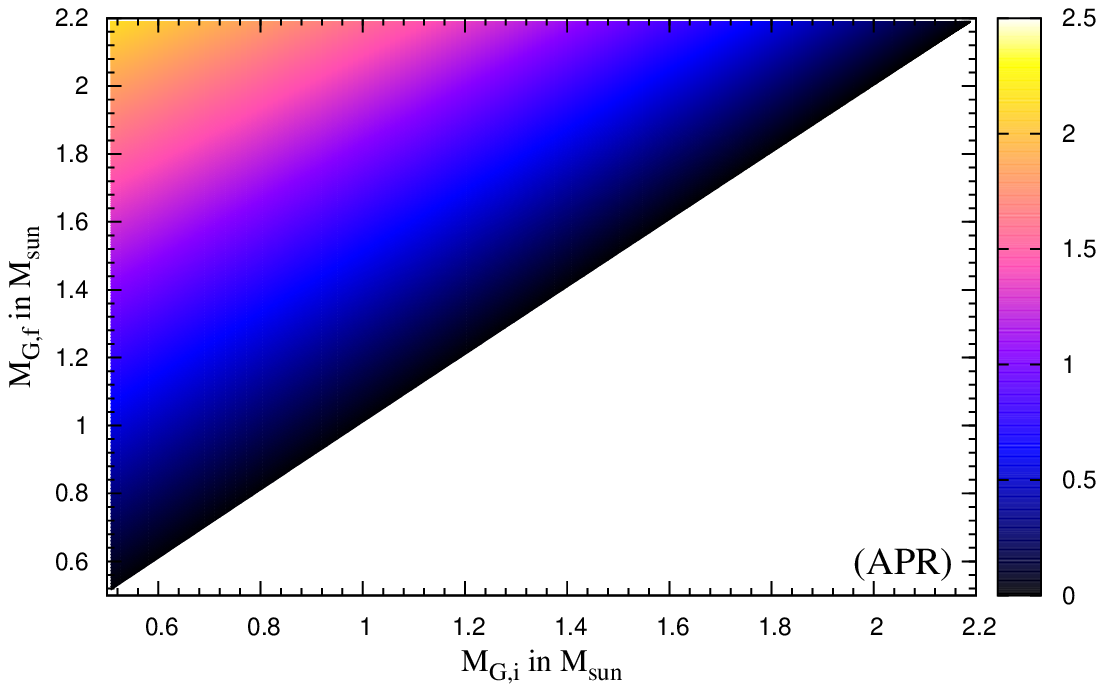,width=10cm}}
\caption{Plot of $M_{G,i}$, $M_{G,f}$ and $M_{acc}$ as obtained using EoS MS0 and EoS APR.} \label{fig:macc}
\end{figure}

In Fig. \ref{fig:macc} we plot $M_{G,i}$ along X-axis, $M_{G,f}$ along Y-axis and $M_{acc}$ as the colour palette. The increase of $M_G$ of a neutron star with the amount of accreted matter can be studied along a straight line parallel to the Y-axis. The X-coordinate of the line will give the initial value of the $M_G$ ($M_{G,i}$) and the Y-coordinate at any point in the line will give the final value of $M_G$ ($M_{G,f}$) for which the amount of mass accreted can be obtained from the colour at that point. But $M_{G,i}$ is not the observable entity, it is $M_{G,f}$ which can be measured. By noting the colours at different points along a straight line parallel to the X-axis, one can get the amount of accreted mass, where the Y-coordinate of the straight line should be set equal to the measured value of $M_{G,f}$ and the X-coordinate of any point along the line will correspond to the assumed value of $M_{G,i}$. Let us substantiate this for the case of the millisecond pulsar PSR J1614-2230 which has $M_{G, f} = (1.97 \pm 0.04) ~M_{\odot}$ \citep{dp10}, spin period 3.15 ms and a companion of mass $0.5~M_{\odot}$. Table \ref{tb:macc} shows the values of $M_{acc}$ for different values of $M_{G, i}$ setting $M_{G, f} = 1.97 ~M_{\odot}$. We find that $M_{acc}$ ($\Delta M_B$) is quite high for $M_{G, i} << M_{G, f}$ which is difficult to explain with present evolutionary models. There are two possibilities to explain the fact, first the neutron star was probably born massive ($M_{G, i} \gtrsim 1.8 ~M_{\odot}$) or the accretion process is much more efficient. Further study is needed to be sure of possible origin of PSR J1614-2230. Moreover, we see from Table \ref{tb:macc} that for fixed values of $M_{G, f}$ and $M_{G, i}$, different EsoS give different values for $M_{acc}$. For a softer EoS, fixed values of $M_{G, f}$ and $M_{G, i}$ give a larger value of $M_{acc}$ than that obtained for a stiffer EoS because of the larger compactness of the neutron star in the first case. Thus proper understanding of binary evolution requires the true knowledge of dense matter EoS. Similarly for any other MSRP, $M_{acc}$ can be estimated with proper choices of $M_{G, i}$ and the EoS. 

\begin{table}
\caption{$M_{acc}$ as obtained using EsoS APR and MS0 for different values of $M_{G, i}$ when $M_{G, f} = 1.97 ~M_{\odot}$ which gives $M_{B, f} = 2.322 ~M_{\odot}$ for EoS APR and $M_{B, f} = 2.237 ~M_{\odot}$ for EoS MS0.}
\begin{center}
\begin{tabular}{cccccc}
\hline
$M_{G, i}$  & $\Delta M_G$   & \multicolumn{2}{c}{$M_{B, i}$} & \multicolumn{2}{c}{$M_{acc} ~(\Delta M_B)$} \\
 ($M_{\odot}$)& ($M_{\odot}$) & \multicolumn{2}{c}{($M_{\odot}$)}  & \multicolumn{2}{c}{($M_{\odot}$)} \\ 
 & & APR & MS0& APR  & MS0   \\
\hline
 1.4 & 0.57 & 1.554 & 1.525 & 0.768 & 0.712\\
 1.5 & 0.47 & 1.681 & 1.647 & 0.641 & 0.591 \\
1.6 & 0.37 & 1.811 & 1.767 & 0.511 & 0.470 \\
1.7 & 0.27 & 1.943 & 1.892 & 0.379  & 0.345 \\
1.8 & 0.17 & 2.080 & 2.018 & 0.242  & 0.219 \\
1.9 & 0.07 & 2.221 & 2.146 & 0.101  & 0.091 \\
 \hline
\end{tabular}
\end{center}
\label{tb:macc}
\end{table}

\section{Discussions} 
\label{sec:discuss}

The extra contribution of the binding energy to the estimate of total accreted mass $M_{acc}$ makes the recycling process more efficient as the conservation of angular momentum gives $I_{f} \, \omega_{f} = I_{i} \, \omega_{i} + M_{acc} \, R_A \, V_A $. Here the angular momentum of the accreted material is given by $J_{acc} =  M_{acc} \, R_A \, V_A $ \citep{kd99}, $R_A$ is the Alfven radius and $V_A$ is the Keplerian velocity at $R_A$; $J_{f} = I_{f} \, \omega_{f}$, $J_{i} = I_{i} \, \omega_{i}$ where $J_{f}$, $I_{f}$, $\omega_{f}$ are the final angular momentum, moment of inertia and angular velocity and $J_{i}$, $I_{i}$, $\omega_{i}$ are the initial angular momentum, moment of inertia and angular velocity. Moment of inertia of a neutron star increases almost linearly with the increase of $M_{G}$ except very close to the maximum mass \citep{mnj10}. So if one perform the calculation considering $M_{acc} = \Delta M_{G} + \Delta BE / c^2$ instead of $M_{acc} = \Delta M_{G}$ with the same choice of all other parameters, $\omega_{f}$ will be higher in the earlier case. The effect of binding energy will be important in any study for $P_{s, f} - M_{G, f}$ correlation as recently done by \citet{zhang10} without considering this effect.

\section{Summary and Conclusions}
\label{sec:conclu}

We find that neutron stars posses significant amount of binding energies which increase with the increase of the gravitational mass of the neutron stars. So when a neutron star gains mass by accreting matter from its companion, its binding energy also increases and the amount of accretion needed becomes greater than the difference in the gravitational masses before and after the accretion process. Indeed it is the difference between the baryonic masses before and after the accretion process. This fact makes it more difficult to explain the evolution of massive MSRPs like PSR J1614-2230. It seems that probably PSR J1614-2230 was born massive ($M_{G, i} \gtrsim 1.8 ~M_{\odot}$). If this is indeed the case, then it is quite possible of discovering more neutron stars with such high gravitational masses - even the neutron stars in the double neutron star systems may have such high gravitational masses whereas in the present sample of double neutron stars, the gravitational masses are usually much smaller - with a mean of $1.35 ~M_{\odot}$ \citep{kkt10}.

We also find that for the same amount of increase in the gravitational mass of a neutron star, different Equations of State predict different values for the increase in the total binding energy. This fact reveals the importance of knowing the true nature of the matter inside a neutron star for better understanding of the evolution of the neutron star through mass accretion from its companion. We use two sample EsoS in this work but the correctness of these are not firmly established. Measurement of the moment of inertia from the faster component of the double pulsar system PSR J0737-3039 \citep{latshuz05} or detection and analysis of low-frequency ($\leqslant 600$ Hz) gravitational waves \citep{ozel10b} might help us to constrain dense matter EsoS in near future.
 
\section*{Acknowledgments}

The author thanks Sushan Konar for discussions and John Miller for a valuable referee report.

\end{document}